\pdfoutput=1

\documentclass[aip,amsmath,amssymb,reprint,nofootinbib,floatfix]{revtex4-1}

\usepackage{graphicx} % Include figure files
\usepackage{dcolumn}% Align table columns on decimal point
\usepackage{bm}% bold math
\usepackage[utf8]{inputenc}
\usepackage[T1]{fontenc}
\usepackage{mathptmx}
\usepackage{etoolbox}

\begin{document}

\title[Electron Emission from Ultracold Plasmas]%
{On the Nature of Subharmonics of the Electron Emission from
Ultracold Plasmas}

\author{Yurii V. Dumin}
\email[Corresponding author's electronic mail: ]{dumin@yahoo.com,\\
dumin@pks.mpg.de}
\affiliation{Space Research Institute of Russian Academy of Sciences, \\
Profsoyuznaya str.\ 84/32, 117997 Moscow, Russia}
\affiliation{Lomonosov Moscow State University,
Sternberg Astronomical Institute, \\
Universitetskii prosp.\ 13, 119234 Moscow, Russia}
\author{Ludmila M. Svirskaya}
\email[Electronic mail: ]{svirskaialm@susu.ru, svirskayalm@mail.ru}
\affiliation{South Ural State University,
Prosp.\ Lenina 76, 454080 Chelyabinsk, Russia}
\affiliation{South Ural State Humanitarian Pedagogical University, \\
Prosp.\ Lenina 69, 454080 Chelyabinsk, Russia}

\begin{abstract}
One of the most interesting phenomena in the ultracold plasmas are multiple
subharmonics of the electron emission observed after its irradiation by
the monochromatic radiowaves.
Unfortunately, the early interpretation of this phenomenon as the so-called
Tonks--Dattner resonances (\textit{i.e.}, actually the standing Langmuir waves)
encountered a number of serious obstacles, such as a lack of the adequate
boundary conditions, an incorrect dependence on the electron temperature,
and an insensitivity to the shape of the cloud.
Here, we suggest an alternative interpretation based on the quasi-classical
multiphoton ionization of the `secondary' Rydberg atoms formed in
the expanding and cooling plasma clouds.
As follows from our numerical simulations, the efficiency of such ionization
exhibits a series of well-expressed peaks.
Moreover, this process is evidently irrelevant to the boundary conditions and
global shape of the cloud.
Therefore, this should be a viable alternative to the earlier idea of
Tonks--Dattner resonances.
\end{abstract}

% \pacs{XX.XX.XX, YY.YY.YY}

\maketitle

\section{Introduction}
\label{sec:Intro}

The physics of ultracold plasmas (UCP) is a relatively new branch of
plasma physics, originated in the very late 1990's and early 2000's due to
advances in the technology of laser cooling of gases and their capture in
the magneto-optical traps\cite{Killian99,Gould01,Bergeson03,Killian07a,%
Killian07b}.
They are the quasi-neutral systems of charged particles with electronic
temperature from a few to several hundreds of Kelvin and the typical ionic
temperature about 1\,K or below.
It is interesting to mention that theoretical studies of such plasmas were
started even before their experimental realization became feasible;
\textit{e.g.}, review\cite{Mayorov94}.
A new impact to the research of UCP was given by the subsequent experiments
on their creation in the continuous regime, both in the supersonic gas-dynamic
jets\cite{Morrison08} and---most recently---by employing the standard
technique of laser cooling\cite{Zelener24}.

The ultracold plasmas were considered initially, first of all, as a promising
tool to achieve considerable values of the Coulomb coupling parameter
(\textit{i.e.}, a ratio of the potential and kinetic energies)\cite{Fortov00}.
However, a number of other interesting and unexpected phenomena were found
in the course of their experimental studies.
One of them are multiple subharmonics of the electron emission observed
after irradiation of UCP by radiowaves.
In other words, when the ultracold plasma bunch expands monotonically and
interacts with a monochromatic electromagnetic wave, the flux of escaping
electrons exhibits a substantially nonmonotonic behavior, with a series of
well-expressed peaks.

In fact, the basic harmonic of this signal was found already in the first
experiments with UCP and became the crucial evidence that a collective
behavior of the electrons was really achieved\cite{Kulin00}.
A few years later, a whole series of the additional peaks was found in
the more elaborated experiment\cite{Fletcher06}; and their interpretation
became challenging.
The first and most evident idea was that they were the subharmonics
of the standing Langmuir waves excited in the plasma cloud of a finite size,
or the so-called Tonks--Dattner~(TD) resonances\cite{Tonks31,Dattner63}.
Namely, it was assumed that the cloud in the course of its expansion
passed through a series of `preferable' radii, which were favorable
for excitation of the standing waves and the resulting electron
outbursts.

Unfortunately, this interpretation encountered a number of serious
obstacles, such as a lack of the adequate boundary conditions in
the freely-expanding plasma cloud without a sharp boundary.
To get around this problem in paper\cite{Fletcher06}, the outer
boundary condition for the plasma waves was imposed at the radius
where the electron Debye length was about the entire size of
the plasma cloud.
Although such an assumption might be reasonable in some approximation,
it is rather crude.
Really, the calculated TD resonances were found to depend appreciably
on the above-mentioned position of the `artificial wall'.
Besides, it was unclear how the high-order TD harmonics---possessing
a nontrivial angular dependence---could be excited by an almost uniform
external electromagnetic field.

Moreover, the pattern of resonances was found to be weakly dependent on
the initial electron energy and, therefore, the subsequent electron
temperature.
This fact was also rather suspicious from the viewpoint of TD paradigm,
because the Bohm--Gross dispersion relation---resulting in the formation
of TD resonances---crucially depends on the electron temperature.
Finally, TD calculations should depend on the shape of the plasma cloud,
while the experimental measurements\cite{Fletcher06} showed that the
subharmonics of the electron emission remained almost the same even in
the strongly distorted clouds, namely, when an opaque wire was placed in
the ionizing laser beam, thereby resulting in the dumbbell-shaped initial
distribution of the plasma density.
(This test was originally employed by the experimentalists to exclude the
ion acoustic waves; but it is actually applicable to any type of the
standing waves.)

Some of the above-listed problems were resolved in the course of
the subsequent studies.
In particular, as follows from the detailed experiment\cite{McQuillen15},
even if a strip of the strongly-depleted density was originally formed
in the plasma cloud, the charged-particle distribution heals over
a few microsecond time scale, and the cloud restores its approximately
Gaussian shape.
So, it might be not so surprising that the electron resonances were
almost insensitive even to the strong initial perturbations.

A considerable theoretical advance in the description of TD resonances
was done in papers\cite{Lyubonko12,Bronin24}.
In particular, employing a dielectric permittivity of the collisional
non-neutral plasmas, a remarkable similarity was found between
the theoretical and experimental resonances; \textit{e.g.}, Fig.~6 in
the last-cited paper\cite{Bronin24}.
Unfortunately, these resonances were actually calculated for the heat
absorbed by the plasma rather than for the escaping electrons.
(In fact, the entire electron motion was assumed to be confined.)
In general, it might be conjectured that the absorbed heat should
stimulate an escape of electrons from the plasma cloud; but this effect
does not follow immediately from the calculations.
Besides, the pattern of resonances remained substantially dependent on
the artificially-introduced cutoff radius of the cloud (the outer
boundary condition); the problem being encountered already in
the earlier study\cite{Fletcher06}.

In view of the above circumstances, it seems reasonable to seek for
the alternative explanations, which should immediately explain the release
of electrons and be independent on the global properties of the plasma
cloud.
It is the aim of the present work to suggest one possible option.
It is based on the radiowave ionization of the `secondary' Rydberg
atoms formed due to recombination in the expanding and cooling plasma
cloud.
Since this approach is purely `local', it completely avoids any
issues associated with boundary conditions and shape of the cloud.
Besides, as will be seen from the subsequent simulations, this
mechanism provides the patterns of peaks of the electron emission
that are qualitatively similar to the observed ones.

\section{Theoretical Model}
\label{sec:Model}

\subsection{Ionization--Recombination Balance in the Expanding
            Plasma Cloud}
\label{sec:Ion-rec_balance}

\begin{figure}
\includegraphics[width=0.65\columnwidth]{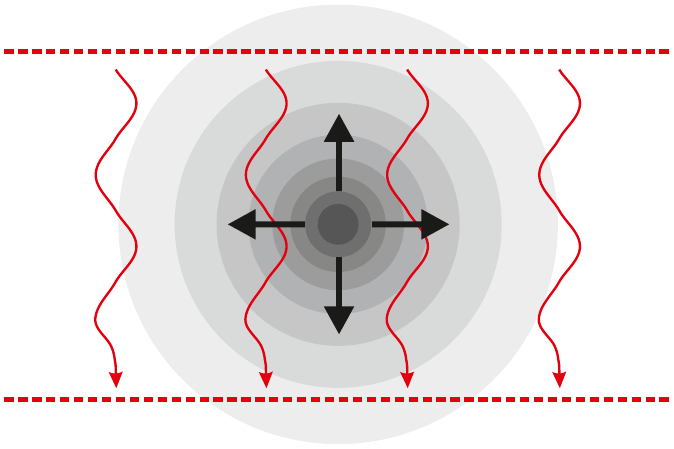}
\caption{\label{fig:Cloud}
Sketch of the plasma cloud expanding inside the rf resonator.}
\end{figure}

As was already mentioned above, we shall consider the ultracold
plasma cloud expanding inside the rf resonator, as illustrated in
Fig.~\ref{fig:Cloud}.
In the course of the expansion and cooling, the plasma decays into
the Rydberg states due to three-body recombination, and
the resulting Rydberg atoms are subsequently ionized again by
the external electromagnetic field.

Assuming for simplicity that the cloud is uniform, the total
flux of escaping electrons can be estimated as
\begin{equation}
I \propto \eta \, \big( dN_{{\rm Ry}} / dt \big)_{\rm \! rec} \, R^3 ,
\label{eq:tot_cur_gen_expr}
\end{equation}
where
$ N_{{\rm Ry}} $~is the concentration of the Rydberg atoms,
$ \eta $~is the efficiency of ionization (\textit{i.e.}, percentage
of the ionized atoms), and
$ R $~is the characteristic radius of the plasma cloud
(for the additional discussion, see Appendix~\ref{sec:Derivation}).

%%%%%%%%%%%%%%%%%%%%%%%%%%%%%%%%%%%%%
%%%%%%%%%%%%%%%%%%%%%%%%%%%%%%%%%%%%%

Next, the rate of collisional three-body recombination is given
by the equation\cite{Massey52}:
\begin{equation}
\bigg( \! \frac{dN_{{\rm Ry}}}{dt} \! \bigg)_{\rm \!\! rec}
  \propto \, N_e^3 \, T_e^{-9/2} ,
\label{eq:three_body_rec}
\end{equation}
where
$ N_e $~is the concentration of the charged particles (\textit{i.e.},
electrons and ions), and
$ T_e $~is the electron temperature (which can be, in general, very
different from the ionic one).
Substituting formula~(\ref{eq:three_body_rec})
into (\ref{eq:tot_cur_gen_expr}), we get the total electron current
in the form:
\begin{equation}
I \propto \eta \, N_e^3 \, T_e^{-9/2} \, R^3 .
\label{eq:tot_cur_partic_expr}
\end{equation}

At the inertial stage of expansion (which actually covers the most
part of evolution of the plasma cloud), $ R(t) \propto t $; so that
\begin{equation}
N_e \propto 1 / R^3 \propto 1 / t^3 .
\label{eq:concentr_time}
\end{equation}
On the other hand, if the processes of inelastic scattering are
assumed to be insignificant%
\footnote{
In fact, the evolution of electron temperature in ultracold plasmas
can deviate substantially from the adiabatic law; \textit{e.g.},
experiment\cite{Fletcher07} and its theoretical
interpretation\cite{Dumin11}.
However, it is hardly reasonable to include such effects into
the present simplified model.
}, then the electron temperature will decay
by the adiabatic law with the adiabatic index $ \gamma = 5/3 $:
\begin{equation}
T_e \propto 1 / R^{3(\gamma - 1)} \propto 1 / R^2 \propto 1 / t^2 .
\label{eq:temperature_time}
\end{equation}

Let us note that all quantities apart from~$ \eta $ in
formula~(\ref{eq:tot_cur_partic_expr}) change monotonically with time.
However, as will be shown in the next section, the efficiency of
ionization~$ \eta $ exhibits a strongly nonmonotonic behavior.
Just this fact results in the observed multiple peaks of the total
electron flux from the plasma cloud.

\subsection{Efficiency of Ionization of Rydberg Atoms}
\label{sec:Ionization}

%%%%%%%%%%%%%%%%%%%%%%%%%%%%%%%%%%%%%
\begin{figure}
\includegraphics[width=0.6\columnwidth]{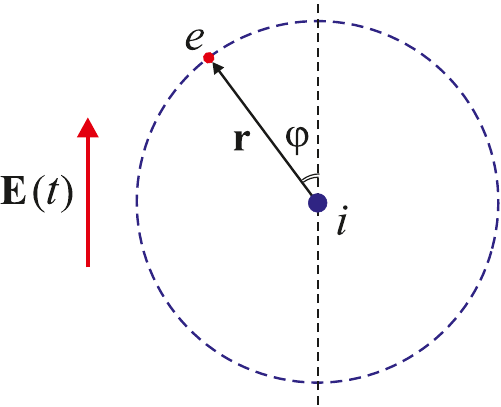}
\caption{\label{fig:Electron}
Motion of the Rydberg electron~($ e $) about the central ion~($ i $)
under the influence of the external electric field~$ \bf E $.}
\end{figure}
%%%%%%%%%%%%%%%%%%%%%%%%%%%%%%%%%%%%%

The most pronounced series of peaks was observed under irradiation
of the cloud by the electromagnetic waves with frequency 20--30\,MHz,
which corresponds to the photon energy~$ 10^{-7} $\,eV; while the
characteristic binding energy of the Rydberg atoms with principal
quantum numbers about~50---which are typically formed after
recombination of ultracold plasmas---should be much greater,
about~$ 5{\times}10^{-3} $\,eV.
Consequently, the ionization proceeds as a successive absorption of
a very large number of photons and can be well described by
the equations of classical mechanics and electrodynamics.
We shall call it the quasi-classical multiphoton ionization.
(Yet another term widely used in the literature---stochastic
ionization---seems to be less appropriate and somewhat misleading,
because such a process can often develop even without onset of
any stochasticity.)
Theoretical studies of this phenomenon were widely conducted
starting from the late 1970's; \textit{e.g.}, review\cite{Delone83}.
The corresponding experiments were performed initially with
the moderately-excited Rydberg atoms ($ n \approx 10{-}30 $) in
the atomic beams irradiated by microwaves\cite{Gallagher94}, so
that such phenomenon was also called the microwave ionization.

The most important finding of the above-mentioned studies---both
theoretical and experimental---was that the efficiency of ionization
exhibits a number of well-expressed peaks, depending on the ratio of
the applied irradiation frequency~$ \omega $ to the frequency of
revolution of the Rydberg electron:
\begin{equation}
\omega / {\Omega}_n =
  \big( {\hbar}^3 / m_e e^4 \big) \, n^3 \, \omega =
  \big( \omega / {\Omega}_1 \big) \, n^3 ,
\label{eq:ratio_freq}
\end{equation}
where
$ \hbar $~is the Planck constant,
$ e $ and $ m_e $~are the electron charge and mass,
$ n $~is the principal quantum number, and
$ {\Omega}_n $~is the `classical' (Bohr) frequency of revolution
of the electron.
The frequency of revolution in the $ n $'th orbit is evidently given
by the relation:
\begin{equation}
{\Omega}_n = \frac{m_e e^4}{{\hbar}^3} \, \frac{1}{n^3} =
  \frac{{\Omega}_1}{n^3} .
\label{eq:Bohr_freq}
\end{equation}
Example of the sharp dependence of the efficiency of ionization
on the ratio~(\ref{eq:ratio_freq}) can be seen, \textit{e.g.},
in Fig.~3 in one of the earliest papers\cite{Jones80}.

%%%%%%%%%%%%%%%%%%%%%%%%%%%%%%%%%%%%%
\begin{figure*}
\includegraphics[width=\textwidth]{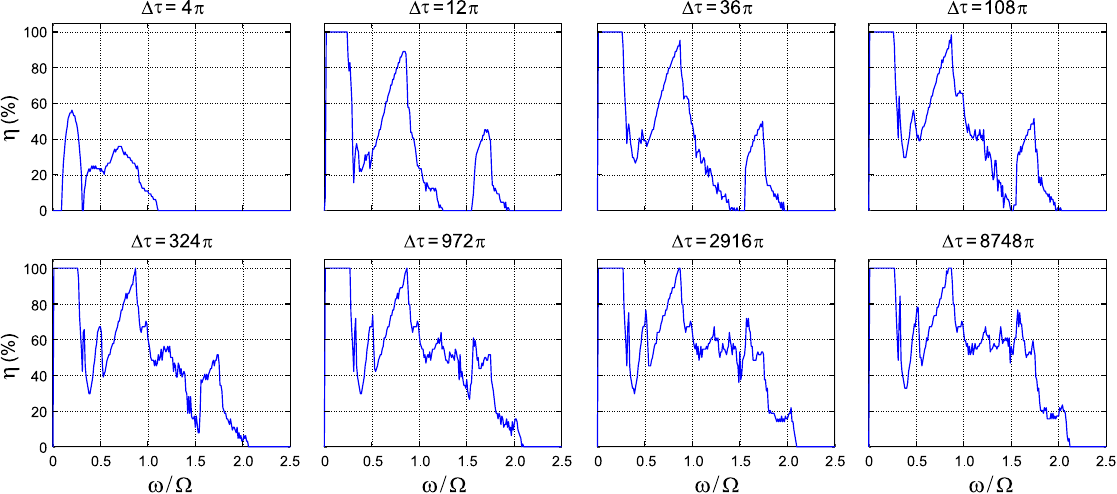}
\caption{\label{fig:Ionization}
Efficiency of ionization~$ \eta $ as function of the ratio of the
electromagnetic wave frequency~$ \omega $ to the frequency of revolution
of the unperturbed Rydberg electron~$ \Omega $.
Different panels correspond to the different duration of interaction of
the Rydberg atom with the electromagnetic wave~$ \Delta \tau $.}
\end{figure*}
%%%%%%%%%%%%%%%%%%%%%%%%%%%%%%%%%%%%%

At last, a well-known property of the three-body
recombination\cite{Massey52} is that it proceeds to the states with
binding energies about the thermal energy of electrons~$ T_e $.
As a result, the corresponding Rydberg atoms are formed in the states
with the principal quantum numbers~$ n^* $:
\begin{equation}
\big| E_{n^*} \big| = \big| E_1 \big| / n^{*2} \propto
  (3/2) \, A \, k \, T_e ,
\label{eq:criter_recomb}
\end{equation}
where
$ k $~is the Boltzmann constant,
$ A $~is a numerical coefficient on the order of unity, and
$ E_n $~is the energy of the $ n $'th atomic level.
In other words,
\begin{equation}
n^* = \bigg( \frac{2 \, | E_1 |}{3 \, A \, k} \bigg)^{\! 1/2} T_e^{-1/2} =
      \bigg( \frac{2 \, | E_1 |}{3 \, A \, k \, T_{e0}} \bigg)^{\! 1/2}
      \big( t / t_0 \big) ,
\label{eq:prin_quant_num}
\end{equation}
\textit{i.e.}, the principal quantum number of the Rydberg atoms
increases linearly with time.
(For simplicity, we assume here that all electrons at the given
temperature~$ T_e $ recombine to the single state~$ n^* $.)

Finally, taking into account
relations~(\ref{eq:concentr_time})--(\ref{eq:ratio_freq}),
formula~(\ref{eq:tot_cur_partic_expr}) for the total electron current
can be rewritten as
\begin{equation}
I \propto \eta (n^*(t)) \, t^3 .
\label{eq:tot_cur_final_expr}
\end{equation}

\subsection{Simulation of the Ionization}
\label{sec:Simulation}

Unfortunately, despite a lot of the earlier results on the efficiency
of quasi-classical multiphoton ionization of Rydberg atoms, it is
impossible to apply them immediately to the problem under consideration.
The case is that such studies usually treated the ionization as
an escape of the electron up to `infinity'.
On the other hand, in the case of plasmas, if the electron is shifted
from the original ion by a distance about a half of the mean
interparticle separation, it begins to experience microfields of other
particles and becomes `collective' (ionized).
So, the ionization of a Rydberg atom in plasmas should be treated as
an escape of its electron up to some threshold radius~$ R_{\rm th} $,
\textit{i.e.}, actually its excitation to a higher orbit.
Just this criterion is the crucial difference of our simulations from
the early ones.
As will be seen below, the overall pattern of the ionization peaks in
such a case can be substantially modified.

The threshold radius can be estimated by the following way.
Because of the above-mentioned property of the three-body recombination,
the recombined atom is formed with a typical binding energy about
the thermal energy of electrons, \textit{i.e.},
\begin{equation}
\frac{e^2}{r_{\rm Ry}} \approx
  \langle K_e \rangle ,
\label{eq:bind_energy}
\end{equation}
where
$ r_{\rm Ry} $~is the characteristic radius of the Rydberg atom, and
$ \langle K_e \rangle $~is the average kinetic energy of an electron.
On the other hand, this kinetic energy can be expressed through
the typical potential energy of the plasma particles as
\begin{equation}
\langle K_e \rangle \approx
  \frac{1}{\Gamma_e} \, \frac{e^2}{\langle r \rangle} ,
\label{eq:relation_energies}
\end{equation}
where
$ \Gamma_e $~is the Coulomb coupling parameter for electrons, and
$ \langle r \rangle $~is the mean separation between the particles
of the same sign.
At last, substituting formula~(\ref{eq:relation_energies}) to
(\ref{eq:bind_energy}), we get the following relation:
\begin{equation}
r_{\rm Ry} \approx \, \Gamma_e \, \langle r \rangle .
\label{eq:Rydberg_size}
\end{equation}

For example, if $ \Gamma_e = 0.2 $, then a typical size of the recombined
Rydberg atom will be about one fifth of the characteristic interparticle
separation.
So, if an electron is located at the distance $ 2.5 \, r_{\rm Ry} $ from
the atomic center, it will experience approximately equal forces from
the two adjacent ions.
Therefore, the threshold radius~$ R_{\rm th} $ should be taken somewhat
below this value.
Particularly, in the most part of our calculations we used
$ R_{\rm th} = 1.5 \, r_{\rm Ry} $.
Anyway, as illustrated in Appendix~\ref{sec:Ioniz-Thresh_rad},
the qualitative pattern of ionization (namely, positions of
the ionization peaks) depends rather weakly on the exact value
of~$ R_{\rm th} $.

Next, the equation of motion of the Rydberg electron perturbed by
the external electric field~$ \bf E $ can be written as
\begin{equation}
m_e \frac{d^2{\bf r}}{d t^2} =
  -e^2 \, \frac{\bf r}{r^3} - e \, {\bf E}(t) ,
\label{eq:electron_motion}
\end{equation}
where $ \bf r $~is the electron radius vector, and the respective
ion is assumed to be localized in the origin of coordinates, as
illustrated in Fig.~\ref{fig:Electron}.

In the polar coordinate system, formula~(\ref{eq:electron_motion}) is
reduced to the set of equations:
\begin{subequations}
\begin{equation}
m_e \big( \ddot{r} - r \, \dot{\varphi}^2 \big) =
  - \frac{e^2}{r^2} -
  e \, E(t) \, \cos \varphi ,
\label{eq:electron_motion_r}
\end{equation}
\begin{equation}
m_e \big( r \, \ddot{\varphi} + 2 \, \dot{r} \, \dot{\varphi} \big) =
  e \, E(t) \, \sin \varphi .
\label{eq:electron_motion_phi}
\end{equation}
\end{subequations}

Since the three-body recombination results in the formation of
Rydberg atoms with large orbital quantum numbers~$ l $ (about the
principal quantum number~$ n $), we shall assume for simplicity that
the initial unperturbed state was purely circular.
So, the unperturbed solution of the above-written equations
(\textit{i.e.}, at $ E \equiv 0 $) will be
\begin{equation}
\varphi = \Omega \, t + {\varphi}_0 , \qquad
\Omega = \big( e^2 \! / m_e \, a_0^3 \big)^{\! 1/2} ,
\label{eq:phi_time_unperturbed}
\end{equation}
where $ a_0 $~is the original unperturbed radius.

Next, let us introduce the dimensionless radius~$ \tilde{r} $ (marked
by a tilde) and time~$ \tau $:
\begin{equation}
r = a_0 \, \tilde{r} , \qquad
t = {\Omega}^{-1} \, \tau ,
\label{eq:dimless_var}
\end{equation}
so that $ \tau = 2 \pi $~is the period of unperturbed revolution.
Substitution of these definitions to the
expressions~(\ref{eq:electron_motion_r})
and~(\ref{eq:electron_motion_phi}) leads us to the dimensionless
equations of motion:
\begin{subequations}
\begin{equation}
m_e a_0 {\Omega}^2 \big( \tilde{r}^{\prime\prime} - \,
  \tilde{r} \, {\varphi}^{\prime 2} \big) =
  - \frac{e^2}{a_0^2} \,
  \frac{1}{\tilde{r}^2} - e \, E(\tau) \, \cos \varphi \, ,
\label{eq:electron_motion_r_dimless}
\end{equation}
\begin{equation}
m_e a_0 {\Omega}^2 \big( \tilde{r} \, {\varphi}^{\prime\prime} + \,
  2 \, \tilde{r}^{\prime} \, {\varphi}^{\prime} \big) =
  e \, E(\tau) \, \sin \varphi \, ,
\label{eq:electron_motion_phi_dimless}
\end{equation}
\end{subequations}
where prime denotes a differentiation with respect to~$ \tau $.

Let the electromagnetic wave be monochromatic, with frequency~$ \omega $
and the initial phase~$ {\psi}_0 $:
\begin{equation}
E(t) = E_0 \, {\sin}\big( \omega \, t + {\psi}_0 \big) =
  E_0 \, {\sin}\Big( \frac{\omega}{\Omega} \tau + {\psi}_0 \Big) \: .
\label{eq:wave_monochrom}
\end{equation}
Then, the set of equations~(\ref{eq:electron_motion_r_dimless})
and~(\ref{eq:electron_motion_phi_dimless}) is reduced to
\begin{subequations}
\begin{equation}
\tilde{r}^{\prime\prime} - \, \tilde{r} \, {\varphi}^{\prime 2} =
  - {\tilde{r}}^{-2} - \,
  \tilde{E}_0 \, {\sin}\big[ ( \omega / \Omega ) \tau + {\psi}^* \big]
  \cos \varphi \, ,
\label{eq:electron_motion_monochrom_r_dimless}
\end{equation}
\begin{equation}
\tilde{r} \, {\varphi}^{\prime\prime} + \,
  2 \, \tilde{r}^{\prime} \, {\varphi}^{\prime} =
  \tilde{E}_0 \, {\sin}\big[ ( \omega / \Omega ) \tau + {\psi}^* \big]
    \sin \varphi \, ,
\label{eq:electron_motion_monochrom_phi_dimless}
\end{equation}
\end{subequations}
where the dimensionless amplitude of the electromagnetic wave is
defined as
\begin{equation}
\tilde{E}_0 = \frac{e E_0}{m_e a_0 {\Omega}^2} \, .
\label{eq:ampl_el_field_dimless}
\end{equation}

%%%%%%%%%%%%%%%%%%%%%%%%%%%%%%%%%%%%%
\begin{figure*}
\includegraphics[width=\textwidth]{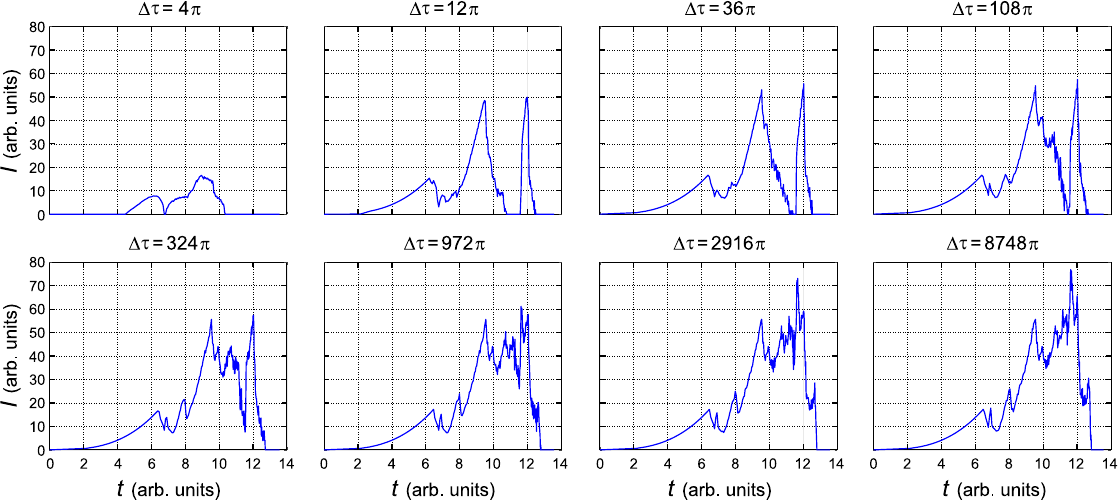}
\caption{\label{fig:Total_current}
Temporal dependences of the total electron current from the expanding
plasma cloud at various durations of interaction of the Rydberg atoms
with the electromagnetic wave.}
\end{figure*}
%%%%%%%%%%%%%%%%%%%%%%%%%%%%%%%%%%%%%

To find the efficiency of ionization~$ \eta $, the set of
equations~(\ref{eq:electron_motion_monochrom_r_dimless})
and~(\ref{eq:electron_motion_monochrom_phi_dimless})
was solved numerically for a sufficiently large sample of the
initial phases (namely, 64~uniformly distributed values).
Then, $ \eta $~was calculated as a percentage of the initial phases
when the computed orbit can reach the threshold radius~$ R_{\rm th} $
(\textit{e.g.}, $ 1.5 \, a_0 $).
As was already explained in the beginning of this section, when
we deal with the ionization in plasmas, it is more reasonable to
consider escape of an electron up to the limited distance rather than
to infinity.

Results of the corresponding simulations are plotted in
Fig.~\ref{fig:Ionization}.
They are presented for various duration of interaction between
the Rydberg atom and the electromagnetic wave~$ \Delta \tau $.
The amplitude of the external electromagnetic wave was taken to
be~$ \tilde{E}_0 = 0.05 $, \textit{i.e.}, 5\%~of the characteristic
interatomic field in the Rydberg atom.
In fact, the subharmonics of electron emission are efficiently formed
within some `optimal' range of the field amplitude
$ \tilde{E}_0 \approx 0.02{-}0.10 $; and we took the value
approximately in the middle of this interval.
(A more detailed discussion of the respective issue can be found in
Appendix~\ref{sec:Ioniz-Field_ampl}.)

For a sufficiently short duration of interaction (\textit{e.g.},
$ \Delta \tau = 4 \pi $), the plot in Fig.~\ref{fig:Ionization} is
rather similar to the early studies\cite{Jones80,Delone83}, where
the detached electron escaped to infinity.
However, at the longer durations~$ \Delta \tau $ one can see
the noticeable differences, namely, the ionization peaks in our
simulations become more diverse and numerous.

At last, substituting the simulated dependences~$ \eta(\omega / \Omega) $
into formula~(\ref{eq:tot_cur_final_expr}) and taking into
account~(\ref{eq:prin_quant_num}), one can get the total electric
current~$ I $ as function of time in the course of the plasma cloud
expansion.
The corresponding plots for various~$ \Delta \tau $ are shown in
Fig.~\ref{fig:Total_current}.
As is seen, a typical pattern of resonances is established starting from
approximately $ \Delta \tau = 12 \pi $ (which corresponds to
$ {\sim}6 $~wave periods), and then it changes rather gradually when
$ \Delta \tau $ increases further.%
\footnote{
Strictly speaking, the dimensionless unit of time was defined in our
simulations in terms of the inverse frequency of the unperturbed
Rydberg electron~$ \Omega $~(\ref{eq:dimless_var}).
However, since all subharmonics are formed at
$ \omega / \Omega \sim 1 $ (within a factor of 2--2.5),
we can assume for the crude estimates that the unit of time is
associated with the inverse wave frequency~$ \omega $.
}

Unfortunately, the characteristic duration of interaction between
the Rydberg atom and the electromagnetic wave cannot be specified
exactly in the framework of our simplified model, where the plasma
cloud was assumed to be uniform, all Rydberg atoms formed in the
same state~(\ref{eq:prin_quant_num}), and the process of rf
ionization occurred almost instantly~(\ref{eq:tot_cur_gen_expr}).
However, we can perform some estimates based on the remark in
paper\cite{Fletcher06} that the patterns of resonances were almost
unchanged when a continuous wave was replaced by the pulses of
1\,$\mu $s duration.
In other words, such a duration was sufficient for the rf ionization
to complete.

%%%%%%%%%%%%%%%%%%%%%%%%%%%%%%%%%%%%%
\begin{figure*}
\includegraphics[width=\textwidth]{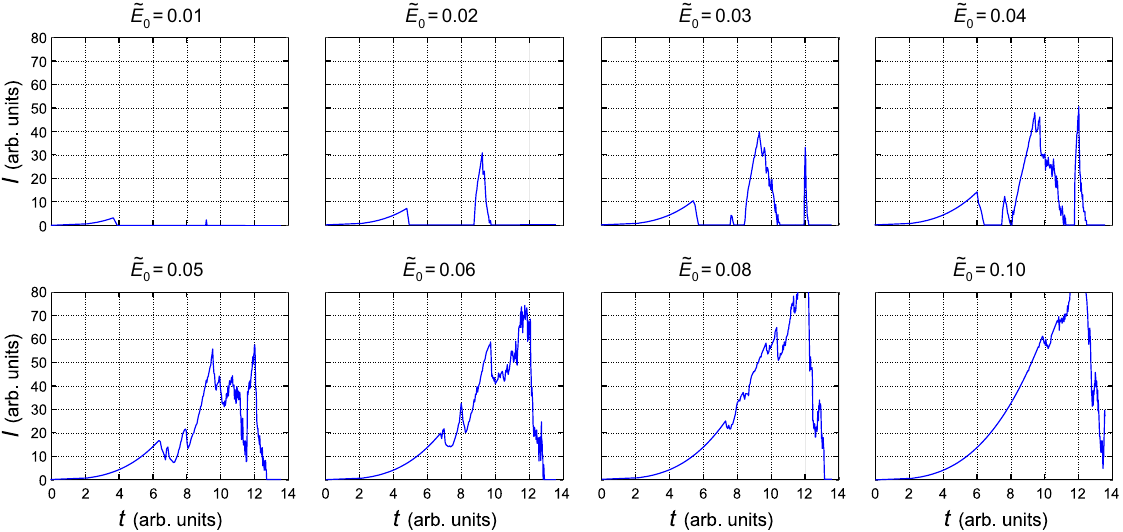}
\caption{\label{fig:Total_current-Field_ampl}
Temporal dependences of the total electron current from the expanding
plasma cloud at various amplitudes of the external electromagnetic
wave~$ \tilde{E}_0 $ (in the dimensionless units, normalized to
the characteristic interatomic field~(\ref{eq:ampl_el_field_dimless})).}
\end{figure*}
%%%%%%%%%%%%%%%%%%%%%%%%%%%%%%%%%%%%%

For example, if we take the characteristic frequency of the external
field $ f = 10$\,MHz (starting from which one can observe
multiple subharmonics in the experiment; see Fig.~1 in
paper\cite{Fletcher06}), then the number of wave periods in the pulse
of duration $ \Delta \, t = 1 $\,$ \mu s $  will be
$ N_{\rm per} = \Delta \, t f \approx 10 $.
This value is in reasonable agreement with the remark in the previous
paragraph that our simulations demonstrate a well-established pattern
of subharmonics starting approximately from 6 wave periods.
Next, if the experimental wave frequency~$ f $ becomes greater, then
the number of periods~$ N_{\rm per} $ within the same pulse
duration~$ \Delta \, t $ will increase.
As is seen in the experimental plots, the resonances in such
conditions become more numerous but closely located and `shallow'.
On the other hand, the simulations in Fig.~\ref{fig:Total_current}
show that, when the number of periods in the pulse
$ N_{\rm per} = \Delta \tau / (2 \pi) $ increases, the subpeaks of
the total current also become more numerous but less distinctive
(\textit{i.e.}, begin to merge to each other).

Therefore, although our simplified model is unable to predict positions
of the resonances as accurately as paper\cite{Bronin24}, the qualitative
behavior of their amplitudes resembles the experimental patterns very
well.

\section{Conclusions}
\label{sec:Conclusions}

In summary, we proposed a new model of formation of subharmonics of
the electron emission from the expanding UCP cloud irradiated by
monochromatic radiowaves.
This model is based on the idea of the quasi-classical multiphoton
ionization of the `secondary' Rydberg atoms formed due to the three-body
recombination.
Our model seems to be a reasonable alternative to the concept of
Tonks--Dattner resonances, because is does not depend on the specific
boundary conditions and is insensitive to the shape of the cloud.
Besides, it naturally explains why the subharmonics of electron
emission become more numerous but less expressed with the increasing
frequency of irradiation.
All these features are in good agreement with the experimental findings.

Of course, a much more detailed \textit{ab initio} simulation should
be performed to draw a definitive conclusion on the viability of this
model.
Such simulation should take into account both nonuniformity
(\textit{e.g.}, a Gaussian density profile) of the cloud and
a nontrivial distribution of the Rydberg atoms over the various
quantum states (as it was done, for example, in paper~\cite{Pohl08}).

\begin{acknowledgments}

We are grateful to J.-M.~Rost for stimulation of this study a few years ago,
as well as to
E.M.~Apfelbaum,
L.G.~Dyachkov,
A.G.~Khrapak,
P.R.~Levashov,
S.A.~Maiorov,
A.D.~Rakhel,
S.A.~Saakyan,
U.~Saalmann,
S.A.~Trigger,
B.B.~Zelener,
B.V.~ Zelener, and
D.I.~Zhukhovitskii
for fruitful discussions.
We are also grateful to the two anonymous referees for a lot of valuable
comments and advices.

\end{acknowledgments}

\section*{Author Declarations}

\subsection*{Conflict of Interest}

The authors have no conflicts to disclose.

\section*{Data Availability}

The data that support the findings of this study are available from the
authors upon reasonable request.

%%%%%%%%%%%%%%%%%%%%%%%%%%%%%%%%%%%%%
\begin{figure*}
\includegraphics[width=\textwidth]{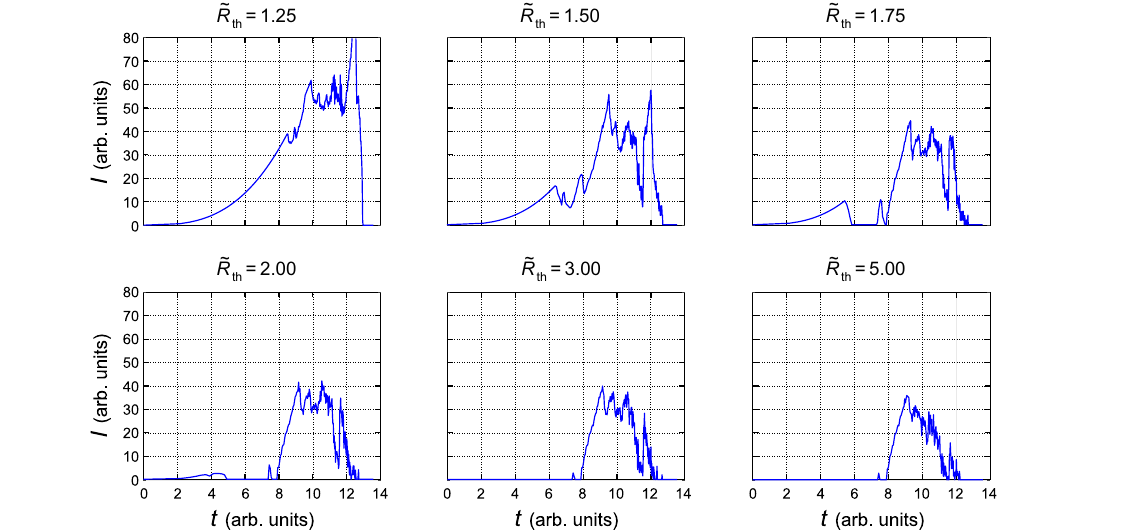}
\caption{\label{fig:Total_current-Thresh_rad}
Temporal dependences of the total electron current from the expanding
plasma cloud for a set of the various threshold
radii~$ \tilde{R}_{\rm th} $, at which the perturbed Rydberg atom is
assumed to be ionized (in the dimensionless units, normalized to the
size of the unperturbed atom~$ a_0 $).}
\end{figure*}
%%%%%%%%%%%%%%%%%%%%%%%%%%%%%%%%%%%%%

\appendix

\section{Derivation of the Expression for the Electric Current}
\label{sec:Derivation}

At the first sight, expression~(\ref{eq:tot_cur_gen_expr}) for the electric
current might look a bit unusual, and a more reasonable choice would be
the more standard formula:
\begin{equation}
I \propto k \, N_{{\rm Ry}} \, R^3 ,
\label{eq:tot_cur_alt_expr}
\end{equation}
where
$ N_{{\rm Ry}} $~is the concentration of Rydberg atoms,
$ R $~is the characteristic radius of the plasma cloud, and
$ k $~is the probability of ionization of an atom in the unit of time.

However, it should be kept in mind that the above-written expression
refers to the case of the ordinary single-photon ionization, when
in each time interval a certain fraction of the available population
of Rydberg atoms is ionized with probability~$ k $, leading to the
exponential decay of that population in the course of time.
But this is not the case of the multi-photon radiofrequency ionization,
when the entire population is affected simultaneously by the
electromagnetic wave.
As a result, some fraction of these atoms (which is denoted by~$ \eta $)
will be quickly ionized within a characteristic time
interval~$ \Delta t $, while the remaining fraction~$ (1 - \eta) $
collapse to the lower energetic states.
So, the initial population will be quickly depleted.
Meanwhile, a new population of the Rydberg atoms---which may have
somewhat different physical properties (\textit{e.g.}, the principal
quantum number)---is formed at the same time interval due to the
recombination.

The above-mentioned process can be described as follows.
The number of free electrons released at the time interval~$ \Delta t $
can be expressed, on the one hand, through the total electron
current~$ I $ and, on the other hand, through the available population
of Rydberg atoms~$ \Delta N_{\rm Ry} (4 \pi / 3) R^3 $ ionized with
the efficiency~$ \eta $:
\begin{equation}
I \, \Delta t\propto \eta \, \Delta N_{\rm Ry} R^3 .
\label{eq:electr_bal}
\end{equation}
Next, dividing both sides of this relation by~$ \Delta t $ and taking
the formal limit~$ \Delta t \to 0 $, we get exactly the
equation~(\ref{eq:tot_cur_gen_expr}).

\section{Dependence of the Ionization Efficiency on the External
         Field Amplitude}
\label{sec:Ioniz-Field_ampl}

In the main body of the present paper, the simulations were
performed at the dimensionless amplitude of the electromagnetic
wave~$ \tilde{E}_0 = 0.05 $ (\textit{i.e.}, 5\%~of the characteristic
interatomic field), which was chosen somewhat arbitrary.
So, it is interesting to check how sensitive are the corresponding
results to the field amplitude.

The answer is given in Fig.~\ref{fig:Total_current-Field_ampl},
which represents simulations at a few different values of
$ \tilde{E}_0 $, while the rf pulse duration is
$ \Delta \tau = 324 \pi $, and all other parameters are the same
as before.
It is seen that separate peaks in the total electron current~$ I $
become visible starting from $ \tilde{E}_0 = 0.02 $.
Next, as the field amplitude increases, the peaks are more numerous
but begin to merge with each other.
As a result, we get the single major peak modulated by a few subpeaks.
When the field amplitude increases further, this modulation becomes
weaker, and finally it looks insignificant at $ \tilde{E}_0 = 0.10 $.
In other words, the subharmonics of the electron emission are
observable in a rather wide range of the rf field amplitudes, from
0.02 to 0.08; so that $ \tilde{E}_0 = 0.05 $ is a quite representative
value.

\section{Dependence of the Ionization Efficiency on the Threshold
         Radius}
\label{sec:Ioniz-Thresh_rad}

Yet another important issue in our simulations is the value of
the threshold radius~$ \tilde{R}_{\rm th} $, starting from which
the electron is assumed to experience appreciable microfields of other
particles and, therefore, can be treated as quasi-free.
In the main body of the paper, we used a rather small value
$ \tilde{R}_{\rm th} = 1.50 $, \textit{i.e.}, only 1.5~times greater
than the radius of an unperturbed Rydberg atom.
So, it is interesting to check how sensitive are our results to
this parameter.

With this aim in view, we performed a series of additional simulations
for various values of~$ \tilde{R}_{\rm th} $, while the parameters
$ \tilde{E}_0 = 0.05 $ and $ \Delta \tau = 324 \pi $ were fixed.
The corresponding results are presented in
Fig.~\ref{fig:Total_current-Thresh_rad}.
It is seen that subharmonics of the total current~$ I $ can be well
identified already at $ \tilde{R}_{\rm th} = 1.25 $.
Then, they become most pronounced in the range of the threshold radii
$ \tilde{R}_{\rm th} = 1.50{-}3.00 $ (apart from the earliest-time
peaks, which gradually disappear) and survive up to
$ \tilde{R}_{\rm th} = 5.00 $.
So, the value $ \tilde{R}_{\rm th} = 1.50 $---used in the most part
of our simulations---gives a rater representative pattern of
subharmonics.

\subsection*{References:}
% \bibliography{Dumin}


\end{document}